\documentclass[runningheads]{llncs}
\usepackage[T1]{fontenc}
\usepackage{graphicx}
\usepackage{multirow}
\usepackage{hyperref}
\usepackage{amsmath}
\usepackage{amsfonts}
\usepackage{float}
\usepackage{siunitx}
\usepackage{xfrac}

\usepackage[utf8]{inputenc} 
\usepackage[T1]{fontenc}    
\usepackage{url}            
\usepackage{booktabs}       
\usepackage{nicefrac}       
\usepackage{microtype}      
\usepackage{xcolor}         
\usepackage{graphicx}
\usepackage{tabularx}

\usepackage{subcaption}

\begin{document}
\title{Image Quality Transfer of Diffusion MRI Guided By High-Resolution Structural MRI}

\titlerunning{IQT of dMRI Guided By High-Resolution Structural MRI}
\author{Alp G. Cicimen\inst{1} \and
Henry F. J. Tregidgo\inst{1} \and
Matteo Figini\inst{1} \and
Eirini Messaritaki\inst{2} \and
Carolyn B. McNabb\inst{2} \and
Marco Palombo\inst{2} \and
C. John Evans\inst{2} \and
Mara Cercignani\inst{2} \and
Derek K. Jones\inst{2} \and
Daniel C. Alexander\inst{1}}
\authorrunning{A. G. Cicimen et al.}
%
\institute{UCL Centre for Medical Image Computing, London, UK \and
Cardiff University Brain Research Imaging Centre, Cardiff, UK
}
\maketitle              
%

%
%
%

\begin{abstract}


Prior work on the Image Quality Transfer on Diffusion MRI (dMRI) has shown significant improvement over traditional interpolation methods. However, the difficulty in obtaining ultra-high resolution Diffusion MRI scans poses a problem in training neural networks to obtain high-resolution dMRI scans. Here we hypothesise that the inclusion of structural MRI\index{Structural MRI} images, which can be acquired at much higher resolutions, can be used as a guide to obtaining a more accurate high-resolution dMRI output. To test our hypothesis, we have constructed a novel framework that incorporates structural MRI scans together with dMRI to obtain high-resolution dMRI scans. We set up tests which evaluate the validity of our claim through various configurations and compare the performance of our approach against a unimodal approach. Our results show that the inclusion of structural MRI scans do lead to an improvement in high-resolution image prediction when T1w data is incorporated into the model input.

\end{abstract}

\section{Introduction}

Image Quality Transfer\index{Image Quality Transfer} by Alexander et al.~\cite{alexander2017image} is a machine learning framework, initially implemented using a random forest (RF) algorithm, that estimates how MRI scans should look if acquired on state-of-the-art scanners. This first implementation showed promise in upsampling dMRI images and has been further developed in works which use convolutional neural networks (CNN)\index{Convolutional Neural Network} to upsample dMRI scans, such as the work by Tanno et al.~\cite{TANNO2021117366}, or on other types of MRI modalities such as the neural network of Lin et al.~\cite{LIN2023102807} or the diffusion model approach of Kim et al.~\cite{kim20233d} for upsampling low-field structural MRI images. While these works do demonstrate the capability of super-resolution (SR)\index{Super Resolution} models to achieve realistic results, they are limited by their upsampling capabilities based on their training configuration. This problem is not easily rectifiable on models that only use dMRI scans as input as obtaining a sufficient Signal-to-Noise Ratio (SNR) at high resolution is a challenging task, especially for dMRI models which require several volumes to be acquired~\cite{JONES2013239}.

Therefore, we propose the incorporation of a secondary MRI modality in conjunction with the low-resolution dMRI input to mitigate the outlined limitations of models that use a single data modality. Prior works, such as the work of Mao et al.~\cite{mao2023discdiff} have demonstrated that a secondary structural MRI modality of different contrast can be incorporated into a model to achieve state-of-the-art results in upsampling structural MRI data. We hypothesise that this incorporation of a second modality can also be used in improving the upsampling capabilities of dMRI In addition, structural MRI scans are easier to obtain at ultra-high resolution and are more readily available, which can help with bypassing the limitations of the dMRI training data. Our approach aims to show as a proof-of-concept that structural MRI scans contain inherent information that a CNN can optimize for and therefore generate better high-resolution dMRI-based computational models.

\section{Methods}

The inputs to our model are a low-resolution dMRI volume and a high-resolution T1w scan. 
We use a similar framework to SynthSR~\cite{IGLESIAS2021118206} in that inputs of varying resolutions are upsampled to a specified target resolution. 
However, instead of processing a whole image volume we use a patch-based network to reduce memory pressure. 
In addition, instead of upsampling specifically to 1mm isotropic resolutions, we implicitly incorporate the desired target resolution in the form of structural information from the T1w companion volume. 
We explain our network architecture and reasoning further in \autoref{sec:Arch}.

For the dMRI-based neural network input, we use the Diffusion Tensor Imaging (DTI)\index{Diffusion Tensor Imaging} model~\cite{BASSER1994259} as it is the simplest and the most commonly used dMRI model.
We specifically use 6 independent elements of the diffusion tensor as inputs.
In \autoref{sec:train_config} and \autoref{sec:eval_config} we explain how we process our input data, and how our training and evaluation configurations differ from each other.

\subsection{Network Architecture}
\label{sec:Arch}

\begin{figure}[t]
    \centering
    \includegraphics[width=\textwidth]{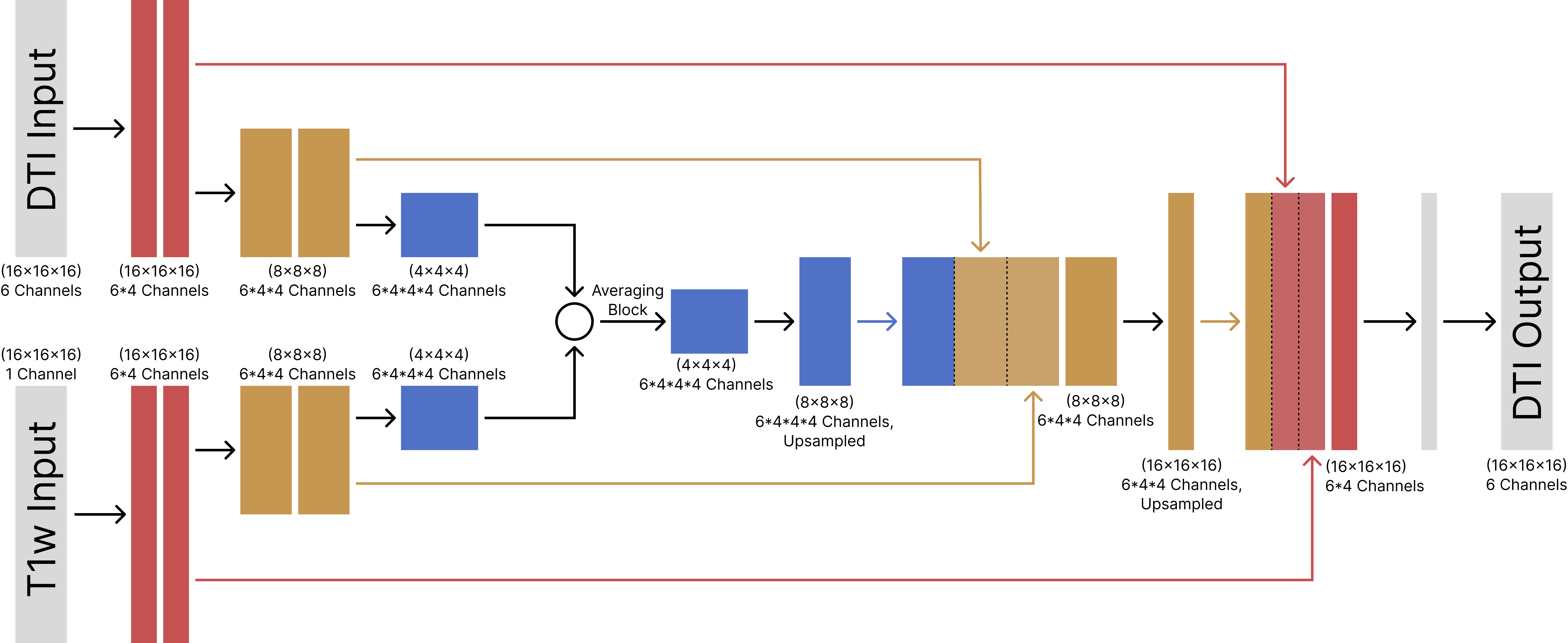}
    \caption{Network diagram for our multimodal UNet architecture designed to test our multimodal approach to super-resolution.
    For this initial feasibility test we limit the network to three levels.}
    \label{fig:model-arch}
\end{figure}

For this network we use a multimodal neural network\index{Multimodal Neural Network} based on a 3D-UNet architecture~\cite{cciccek20163d}\index{UNet}. 
In addition to being a generally versatile model, similar approaches in applying super resolution~\cite{LIN2023102807,IGLESIAS2021118206} have demonstrated the capability of UNet-like models in generating high-resolution images. 
Due to previous works outlining better performance at capturing common features~\cite{ngiam2011multimodal,boulahia2021early,steyaert2023multimodal}, we fuse information from our dMRI and T1w images at the latent layer. 
We have modified the UNet architecture by using a secondary encoder for the T1w input, using a simple averaging operation to join the two latent encodings together. 
For each encoder and decoder layer, we have repeating convolutions without residual connections to further capture possible information. 
We include skip connections between the decoder layers and layers from both encoders at the corresponding level, allowing transfer of high-level information from both modalities. 
A visual diagram of our model can also be found in \autoref{fig:model-arch}.

Our model inputs are a high-resolution T1w patch $\mathcal{P}^s_{HR}$ of shape $(16\times16\times16\times1)$ and a DTI low-resolution patch $\mathcal{P}^d_{LR}$ of shape $(16\times16\times16\times6)$ and outputs a high-resolution DTI patch $\mathcal{P}^d_{HR}$ of shape $(16\times16\times16\times6)$. 
Here we use a pre-upsampling approach, where the input dMRI image is initially upsampled linearly to the resolution of patch $\mathcal{P}^s_{HR}$ to generate patch $\mathcal{P}^d_{LR}$ before both are passed into the model. 
This allows a dynamic upsampling rate to the T1w resolution which would not be possible with a post-upsampling model, where the model has a preconfigured upsampling rate.

In addition to our fusion model, we use a unimodal UNet for comparison in ablation experiments. 
This is a standard, single-encoder UNet architecture that only uses the DTI as input. 
We have configured our base model such that the number of hidden layers and channels allocated to the DTI encoder is the same as the channels allocated in the DTI only ablation encoder.
This means any increase in performance can be attributed to the T1w weights rather than an increase in available weights for the DTI input.

\subsection{Training Configuration}
\label{sec:train_config}

To train our model we need to supply triplets of patches, $\mathcal{P}^D_{HR}$, $\mathcal{P}^D_{LR}$ and $\mathcal{P}^S_{HR}$, ensuring these three patches are sampled on the same discretisation. 
As the T1w images in our training set are higher resolution than the available high-resolution DTI, we need to downsample our T1w images while correcting for aliasing effects. 
For this, we use the same approach as SynthSeg~\cite{billot2020learning,10.1007/978-3-030-59728-3_18} to smooth the input T1w with an appropriate Gaussian kernel for the target resolution before downsampling. 

Our input low resolution DTI is generated using a similar process, starting by Gaussian blurring and linearly downsampling to our low-resolution space from the original DTI data. 
We then linearly upsample the low-resolution DTI to the target discretisation of our T1w input. 
This results in images with voxel spacing at the target resolution, where the voxels are the same shape as our target image, but have a lower effective resolution. 
Our base model uses multiple effective $\mathcal{P}^D_{LR}$ resolutions during the training process to generalise performance with multiple low-resolution DTI configurations. 
Our training and testing resolution configurations can be found in detail in \autoref{sec:Experiments}.

After the resampling steps, we clip our data to a specific range and apply min-max normalisation to both the DTI and T1w inputs. 
For DTI we normalise and clip both the low-resolution and high-resolution images to the range $[0,~2\times10^{-3}]$~\si{mm^2s^{-1}} for the diagonal elements $D_{xx}$, $D_{yy}$ and $D_{zz}$, and $[-2\times10^{-3},~2\times10^{-3}]$~\si{mm^2s^{-1}}  for the non-diagonal elements $D_{xy}$, $D_{xz}$ and $D_{yz}$ respectively. 
We have found that these clipping values preserve the information in the non-CSF regions, while dynamic clipping can excessively compress values in these regions. 
For the T1w input we clip and normalise our image to between the 2nd and 98th percentile of masked voxel intensities in our training set.

The HR DTI, HR T1w and LR DTI scans for each training subject are then split into patches of size $16\times16\times16$ which we concatenate into a patch triplet of shape $(16, 16, 16, 13)$. 
During training, we add random augmentations in the form of noise, brightness adjustments and gamma scaling to the pre-processed T1w input to ensure robustness of the model.

\subsection{Evaluation and Inference Configuration}
\label{sec:eval_config}

To evaluate our model we use similar preprocessing steps as in training to generate input patch pairs. 
However, patches are selected such that neighbouring patches overlap by four voxels in each dimension. 
To recombine patches into a full volume we blend patch edges by averaging the overlapping regions resulting in a better image output quality. 

After the model outputs are reconstructed into a single image, we scale back the output images to the original DTI scale using the clipped parameters outlined in \autoref{sec:train_config}. 
The resulting data can now be evaluated as is, or with DTI-based metrics as explained in \autoref{sec:Experiments}.

\section{Experiments}
\label{sec:Experiments}

This section presents the experimentation procedure that we used to evaluate the performance of our model. For each setup we tested in isolation a specific aspect of our model against an ablated one, which was configured as either a unimodal network or a model that used a single training input resolution. 

\subsection{Experiment Data}
\label{sec:Data}

We use scans from two different data sources for our quantitative and qualitative experiments. For our quantitative experiments, we use the datasets in the WU-Minn consortium from the Human Connectome Project~\cite{VANESSEN20122222} for training and testing purposes. The consortium is a multi-institutional collaboration that contains multiple datasets consisting of MRI and DWI sequences obtained using a customized Siemens 3T scanner with a gradient field strength of $300\text{mT}/\text{m}$. We use 174 subjects from the HCP Young Adult dataset, from which we specifically use the T1w scan with 0.7mm isotropic voxel spacing, as well as a DWI sequence that consists of 90 diffusion-weighted images with b-values of 1000$\text{s/mm}^2$, with a voxel size of 1.25mm. The sequence contains 18 $b=0$ scans as well, resulting in a total of 108 dMRI scans and 1 T1w image per subject that we use. We calculate our model inputs by fitting the DTI model to the dMRI images. We split our data on a per-subject base into three parts, which act as our training, validation and testing sets with a split percentage of 70\%, 10\% and 20\% respectively.

For our qualitative results, we use a second data source for evaluating the network that has been trained on the HCP dataset. For this purpose, we use an out-of-distribution scan obtained through two different MRI scanners. Our DWI sequence was obtained through a Connectom 3T scanner using gradient field strengths of 300$\text{mT}/\text{m}$, while our structural scan was obtained through a 7T scanner. While the specifications of the Connectom scanner are the same as the scanner used for obtaining the dMRI scans in the HCP dataset, our DWI sequence is limited in terms of image resolution of 2mm isotropic voxel size, and it consists of 53 images with a maximum b-value of $b=\text{1200s/mm}^2$.

\subsection{Experiment Setup}

\begin{table}[t]
\centering
    \resizebox{\textwidth}{!}{
        \begin{tabular}{|c|c||c|c|c|c|c|}
            \hline
            \begin{tabular}[c]{@{}c@{}}Test\\ Name\end{tabular}
            & \begin{tabular}[c]{@{}c@{}}Model\\ Configuration\end{tabular}
            & \begin{tabular}[c]{@{}c@{}}T1w\\ Included?\end{tabular}
            & \begin{tabular}[c]{@{}c@{}}Training DTI Input\\ Resolution(s) (mm)\end{tabular} 
            & \begin{tabular}[c]{@{}c@{}}Training\\ Target\\ Resolution (mm)\end{tabular}
            & \begin{tabular}[c]{@{}c@{}}Testing DTI Input\\ Resolution(s) (mm)\end{tabular}
            & \begin{tabular}[c]{@{}c@{}}Testing\\ Target\\ Resolution (mm)\end{tabular}\\ \hline\hline
            \multirow{7}{*}{1: Baseline}
            & Standard
            & Yes 
            & \begin{tabular}[c]{@{}c@{}}\{1.5625, 1.875, 2.5, 3.125\}\end{tabular} 
            & \multirow{7}{*}{1.25}
            & \multirow{7}{*}{\begin{tabular}[c]{@{}c@{}}\{1.875, 3.125\}\end{tabular}}
            & \multirow{7}{*}{\begin{tabular}[c]{@{}c@{}}1.25\end{tabular}}\\ \cline{2-3}\cline{4-4}
            & \begin{tabular}[c]{@{}c@{}}Single Resolution\\ $1.5625$mm\end{tabular}
            & Yes
            & \begin{tabular}[c]{@{}c@{}}1.5625\end{tabular}
            &
            &
            & \\ \cline{2-3}\cline{4-4}
            & \begin{tabular}[c]{@{}c@{}}Single Resolution\\ $3.125$mm\end{tabular}
            & Yes
            & \begin{tabular}[c]{@{}c@{}}3.125\end{tabular}
            &
            &
            & \\ \cline{2-3}\cline{4-4}
            & \begin{tabular}[c]{@{}c@{}}Multiple Resolution\\ DTI only\end{tabular}
            & No
            & \begin{tabular}[c]{@{}c@{}}\{1.5625, 1.875, 2.5, 3.125\}\end{tabular}
            &
            &
            & \\ \hline
            \multirow{7}{*}{\begin{tabular}[c]{@{}c@{}}2: OOD\\ Upsampling Rate\end{tabular}}
            & Standard
            & Yes
            & \begin{tabular}[c]{@{}c@{}}\{1.5625, 1.875, 2.5, 3.125\}\end{tabular}
            & \multirow{7}{*}{1.25}
            & \multirow{7}{*}{3.75}
            & \multirow{7}{*}{1.25}\\ \cline{2-3}\cline{4-4}
            & \begin{tabular}[c]{@{}c@{}}Single Resolution\\ $1.5625$mm\end{tabular}
            & Yes
            & \begin{tabular}[c]{@{}c@{}}1.5625\end{tabular}
            &
            &
            & \\ \cline{2-3}\cline{4-4}
            & \begin{tabular}[c]{@{}c@{}}Single Resolution\\ $3.125$mm\end{tabular}
            & Yes
            & 3.125
            &
            &
            & \\ \cline{2-3}\cline{4-4}
            & \begin{tabular}[c]{@{}c@{}}Multiple Resolution\\ DTI only\end{tabular}
            & No
            & \begin{tabular}[c]{@{}c@{}}\{1.5625, 1.875, 2.5, 3.125\}\end{tabular}
            &
            &
            & \\ \hline
            \multirow{4}{*}{\begin{tabular}[c]{@{}c@{}}3: OOD\\ Target Resolution\end{tabular}}
            & Standard
            & Yes
            & \begin{tabular}[c]{@{}c@{}}\{2.8, 3.0, 3.2, 3.4\}\end{tabular}
            & \multirow{4}{*}{2.0}
            & \multirow{4}{*}{2.0}
            & \multirow{4}{*}{1.25} \\ \cline{2-3}\cline{4-4}
            & \begin{tabular}[c]{@{}c@{}}Single Upsample Rate\end{tabular}
            & Yes
            & \begin{tabular}[c]{@{}c@{}}3.2\end{tabular}
            & 
            & 
            & \\ \cline{2-3}\cline{4-4}
            & \begin{tabular}[c]{@{}c@{}}Single Upsample Rate\\ DTI only\end{tabular}
            & No
            & \begin{tabular}[c]{@{}c@{}}3.2\end{tabular}
            & 
            & 
            & \\ \hline
            \multirow{3}{*}{\begin{tabular}[l]{@{}l@{}}4: Qualitative\end{tabular}}
            & Standard
            & Yes
            & \begin{tabular}[c]{@{}c@{}}\{1.5625, 1.875, 2.5, 3.125\}\end{tabular}
            & \multirow{3}{*}{1.25}
            & \multirow{3}{*}{2.0}
            & \multirow{3}{*}{0.7} \\ \cline{2-3}\cline{4-4}
            & \begin{tabular}[c]{@{}c@{}}Single Upsample Rate\\ DTI only\end{tabular}
            & No
            & \begin{tabular}[c]{@{}c@{}}3.75\end{tabular}
            & 
            & 
            & \\ \hline
        \end{tabular}
    }
\caption{Description of network parameters for the four experiments. 
Here we list the isotropic input and target voxel dimensions used in training and testing.
Grouped resolutions indicate the network was trained or tested on multiple upsampling rates.}
\label{tab:experiment-config}
\end{table}

To evaluate the model performance, we have devised 4 tests, with the configurations described in \autoref{tab:experiment-config}, that each individually evaluate in-distribution performance, performance on unseen upsampling factors, and performance when targeting unseen higher resolutions. The 4 tests were devised as follows:

\vspace{2mm}
\noindent\textbf{Test 1: Baseline} Our initial test compares the baseline performance of the multimodal approach to three ablation models. 
For this test, we train our neural network on multiple low-resolution DTI inputs to upsample to our target resolution. 
The model is then evaluated on 2 in-distribution upsampling rates, a $1.5\times$ upsampling rate (from 1.875 to 1.25~\si{mm} voxel dimensions) and a $2.5\times$ upsampling rate (from 3.125 to 1.25~\si{mm} voxel dimensions). 
This tests the general performance of the multi-resolution training against corresponding single-resolution networks. 
In addition, we compare the model's performance to an ablation model that uses DTI input but no T1w input.

\vspace{2mm}
\noindent\textbf{Test 2: Out of Distribution Upsampling} 
Here we test our models on upsampling an out-of-distribution 3.75mm DTI image input to evaluate how the models in Test 1 behave on previously unseen upsampling rates. 
This tests if the addition of the T1w image compensates for the lost details in further downsampled DTI image. 
A higher performance of the multimodal network over our unimodal ablation model would demonstrate that the T1w image contains useful information for upsampling DTI data.

\vspace{2mm}
\noindent\textbf{Test 3: Out of Distribution Target Resolution} For our final quantitative test we evaluate whether we can achieve resolutions higher than those available in the training set. 
We train our network on upsampling DTI images of multiple resolutions to a single target resolution $\mathcal{R}'_{train}$, selected such that $\mathcal{R}'_{train}$ is worse than the test resolution $\mathcal{R}'_{test}$. 
For this test, our ablation models differ by using only a single upsampling rate $\mathcal{U}_{ablation} = \sfrac{\mathcal{R}'_{train}}{\mathcal{R}'_{test}}$. 
This is because there isn't an explicit method to use a dynamic upsampling rate without the use of extra parameters.
Here an improved performance of the multimodal network over the unimodal model would demonstrate its ability to infer information from the T1w image and use it to correct features not discernible from lower quality DTI input.

\vspace{2mm}
\noindent\textbf{Test 4: Qualitative} In addition to the quantitative tests we also include a result for the qualitative output of our model compared to an unimodal ablation model trained on a $3\times$ upsampling rate. We use the Coloured FA map to visualise how the addition of T1w input from a 7T scanner affects the high-resolution model output.

\vspace{2mm}
To obtain our quantitative results, we used 34 test subjects from the HCP dataset that were selected at random. We evaluate the performance of each model using the metrics described in \autoref{sec:Evaluation} and evaluate the statistical significance of of differences using the Wilcoxon's Signed Rank test with a significance threshold of 5\%. Our qualitative result was obtained from a single test subject acquired as outlined in \autoref{sec:Data}.

We have trained our models for 100 epochs, with 400 patches selected per subject per epoch. We use a minibatch size of 40 per iteration and use the L1 loss through the batch as our loss function. For optimisation, we have used Adam by Kingma et al.~\cite{kingma2017adam} and an initial learning rate of $1\times10^{-3}$, with an exponential learning rate decay that halves every 10 epochs.

\subsection{Experiment Evaluation}
\label{sec:Evaluation}

To evaluate our model performance we have used both the DTI output directly and its derived metrics. We initially evaluate the model performance on the DT-RMSE metric defined by Alexander et al.~\cite{alexander2017image} as 
\begin{equation}
    \textrm{median}_{v\in\Omega(i)}\left(\sqrt{\frac{1}{6}\sum_{j=1}^{6}(D_j-D^{*}_j)^2}\right)
\label{eq:DT-RMSE}
\end{equation}
where $D_j$ and $D^*_j$ are the predicted and ground truth j-th DTI element for every voxel $v$ that is contained within the set of all masked regions in a single subject $\Omega(i)$. For the derived metrics we consider the Mean Diffusivity and Fractional Anisotropy as defined by Basser et al.~\cite{basser2011microstructural} and evaluate the Root-Mean-Squared Error (RMSE) as well as the Structural Similarity Index Measure (SSIM)~\cite{995823}, which assesses the general visual performance of our model on the aforementioned metrics. In addition, we investigate the Coloured FA maps~\cite{pajevic2000color} using the mean cosine similarity (CSIM) between the predicted and target vectors, defined as
\begin{equation}
    \text{CSIM}(\mathbf{x}, \mathbf{x}^*)  = \frac{1}{|\Omega|}\sum_{i\in\Omega}\left(\frac{|\mathbf{x}_i^* \cdot \mathbf{x}_i|}{||\mathbf{x}_i^*||\cdot||\mathbf{x}_i||}\right)
\label{eq:CosSim}
\end{equation}

\section{Results}

The quantitative results of our experiments can be viewed in \autoref{tab:results}.

\begin{table}[tb]
    \centering
    \resizebox{\textwidth}{!}{
        \begin{tabular}{|c|c||c|c|c||c|c|c|}
            \hline
            \begin{tabular}[c]{@{}c@{}}Test\\ Name\end{tabular}
            &\begin{tabular}[c]{@{}c@{}}Model\\ Configuration\end{tabular}
            &DT - RMSE$\downarrow$&MD RMSE$\downarrow$&FA RMSE$\downarrow$&MD SSIM$\uparrow$&FA SSIM$\uparrow$&CFA CSIM$\uparrow$\\ \hline\hline
            
            \multirow{9}{*}{\begin{tabular}[c]{@{}c@{}}1: Baseline\\1.875mm Input\end{tabular}}
            &\begin{tabular}[c]{@{}c@{}}Linear\\Interpolation\end{tabular}
            &$6.711\times10^{-5}$&$1.463\times10^{-4}$&$1.005\times10^{-1}$&$0.962$&$0.943$&$0.930$\\ \cline{2-8}
            &\begin{tabular}[c]{@{}c@{}}Multiple Resolution\\Standard Model\end{tabular}
            &$3.489\times10^{-5}$&$6.060\times10^{-5}$&$4.537\times10^{-2}$&$0.993$&$0.986$&$0.963$\\ \cline{2-8}
            &\begin{tabular}[c]{@{}c@{}}Single Resolution\\$1.875$mm\end{tabular}
            &$\underline{\mathbf{3.337\times10^{-5}}}$&$\underline{\mathbf{5.794\times10^{-5}}}$&$\underline{\mathbf{4.504\times10^{-2}}}$&$\underline{\mathbf{0.994}}$&$\underline{\mathbf{0.987}}$&$\underline{\mathbf{0.964}}$\\ \cline{2-8}
            &\begin{tabular}[c]{@{}c@{}}Single Resolution\\ $3.125$mm\end{tabular}
            &$1.427\times10^{-4}$&$1.725\times10^{-4}$&$1.593\times10^{-1}$&$0.965$&$0.944$&$0.939$\\ \cline{2-8}
            &\begin{tabular}[c]{@{}c@{}}Multiple Resolution\\No T1w Input\end{tabular}
            &$3.568\times10^{-5}$&$6.744\times10^{-5}$&$4.767\times10^{-2}$&$0.992$&$0.986$&$0.962$\\ \hline\hline

            \multirow{9}{*}{\begin{tabular}[c]{@{}c@{}}1: Baseline\\3.125mm Input\end{tabular}}
            &\begin{tabular}[c]{@{}c@{}}Linear\\Interpolation\end{tabular}
            &$1.028\times10^{-4}$&$2.068\times10^{-4}$&$1.504\times10^{-1}$&$0.923$&$0.879$&$0.890$\\ \cline{2-8}
            &\begin{tabular}[c]{@{}c@{}}Multiple Resolution\\Standard Model\end{tabular}
            &$4.701\times10^{-5}$&$\underline{\mathbf{8.725\times10^{-5}}}$&$5.966\times10^{-2}$&$\underline{\mathbf{0.985}}$&$\underline{0.976}$&$0.946$\\ \cline{2-8}
            &\begin{tabular}[c]{@{}c@{}}Single Resolution\\ $1.875$mm\end{tabular}
            &$7.803\times10^{-5}$&$1.810\times10^{-4}$&$1.091\times10^{-1}$&$0.943$&$0.935$&$0.911$\\ \cline{2-8}
            &\begin{tabular}[c]{@{}c@{}}Single Resolution\\ $3.125$mm\end{tabular}
            &$\underline{\mathbf{4.624\times10^{-5}}}$&$9.474\times10^{-5}$&$\underline{5.938\times10^{-2}}$&$0.983$&$0.974$&$\underline{\mathbf{0.948}}$\\ \cline{2-8}
            &\begin{tabular}[c]{@{}c@{}}Multiple Resolution\\No T1w Input\end{tabular}
            &$5.277\times10^{-5}$&$1.170\times10^{-4}$&$6.736\times10^{-2}$&$0.977$&$0.968$&$0.940$\\ \hline\hline

            \multirow{9}{*}{\begin{tabular}[c]{@{}c@{}}2: Out of Dist.\\Upsampling Rate\end{tabular}}
            &\begin{tabular}[c]{@{}c@{}}Linear\\Interpolation\end{tabular}
            &$1.141\times10^{-4}$&$2.259\times10^{-4}$&$1.734\times10^{-1}$&$0.912$&$0.857$&$0.871$\\ \cline{2-8}
            &\begin{tabular}[c]{@{}c@{}}Multiple Resolution\\Standard Model\end{tabular}
            &$7.795\times10^{-5}$&$1.557\times10^{-4}$&$1.168\times10^{-1}$&$0.955$&$0.931$&$0.911$\\ \cline{2-8}
            &\begin{tabular}[c]{@{}c@{}}Single Resolution\\ $1.875$mm\end{tabular}
            &$1.032\times10^{-4}$&$2.147\times10^{-4}$&$1.522\times10^{-1}$&$0.921$&$0.888$&$0.882$\\ \cline{2-8}
            &\begin{tabular}[c]{@{}c@{}}Single Resolution\\ $3.125$mm\end{tabular}
            &$\underline{\mathbf{5.802\times10^{-5}}}$&$\underline{\mathbf{1.195\times10^{-4}}}$&$\underline{\mathbf{8.015\times10^{-2}}}$&$\underline{\mathbf{0.974}}$&$\underline{\mathbf{0.959}}$&$\underline{\mathbf{0.932}}$\\ \cline{2-8}
            &\begin{tabular}[c]{@{}c@{}}Multiple Resolution\\No T1w Input\end{tabular}
            &$8.617\times10^{-5}$&$1.788\times10^{-4}$&$1.265\times10^{-1}$&$0.940$&$0.911$&$0.897$\\ \hline\hline


            \multirow{7}{*}{\begin{tabular}[c]{@{}c@{}}3: Out of Dist.\\Target Resolution\end{tabular}}
            &\begin{tabular}[c]{@{}c@{}}Linear\\Interpolation\end{tabular}
            &$7.222\times10^{-5}$&$1.612\times10^{-4}$&$1.053\times10^{-1}$&$0.953$&$0.930$&$0.926$\\ \cline{2-8}
            &\begin{tabular}[c]{@{}c@{}}Multiple Resolution\\Standard Model\end{tabular}
            &$5.280\times10^{-5}$&$1.060\times10^{-4}$&$7.506\times10^{-2}$&$0.976$&$0.962$&$0.943$\\ \cline{2-8}
            &\begin{tabular}[c]{@{}c@{}}With T1w Input\\ $1.6\times$ Upsampling Rate\end{tabular}
            &$\underline{\mathbf{4.982\times10^{-5}}}$&$\underline{\mathbf{9.798\times10^{-5}}}$&$\underline{\mathbf{6.759\times10^{-2}}}$&$\underline{\mathbf{0.980}}$&$\underline{\mathbf{0.968}}$&$\underline{\mathbf{0.945}}$\\ \cline{2-8}
            &\begin{tabular}[c]{@{}c@{}}Without T1w Input\\ $1.6\times$ Upsampling Rate\end{tabular}
            &$5.563\times10^{-5}$&$1.058\times10^{-4}$&$8.407\times10^{-2}$&$0.975$&$0.960$&$0.936$\\ \hline
        \end{tabular}
    }
    \caption{Median RMSE and SSIM values across test subjects for experiments one to three. 
    For each experiment, the best performing model is underlined and highlighted in bold when differences to the second best model reach significance determined by a Wilcoxon's signed rank test.}
    \label{tab:results}
\end{table}

Our baseline in-distribution results demonstrate that the incorporation of a T1w input does improve the model's prediction capabilities. The difference between the result metrics of our multiple resolution approach and our unimodal ablation model was statistically significant, with the median values for the errors being lower and the similarity metrics being higher for our multimodal approach. However, for $1.875\text{mm}$ input resolution we noticed that the single resolution ablation model trained at its native input resolution performed the best out of all inputs. We however have noticed that the aforementioned model tended to break down at lower resolutions as evident in the $3.125\text{mm}$ input configuration. We have also noticed a similar behaviour from the $3.125\text{mm}$ input ablation model when it was provided with the $1.875\text{mm}$ input. Compared to both models our multiple resolution approach demonstrates robustness in both configurations compared to a single resolution training approach.

Our out-of-distribution results for a higher upsampling rate have also demonstrated that the addition of a T1w input does improve the model output. We have observed a difference between our multimodal approach and unimodal approach that was statistically significant, for which we observed that every metric demonstrated a preference for the T1w model. However, the results also demonstrated a preference for the single resolution $3.125\text{mm}$ input model compared to the multiple resolution model.

The results that we have obtained demonstrate differences in quality across our main model compared to our ablation models. In most of our experiments, we have observed an increase in upsampling quality on models that incorporated the T1w images compared to the unimodal ablation model. The ablation models that were evaluated in their native resolution usually performed the best at specified resolutions. Our multi-resolution training configuration with the inclusion of a high-resolution MRI modality achieved comparable results to the ablation models on higher quality DTI inputs and demonstrated better performance when the quality of the DTI inputs was lower ($>2mm$ voxel size). 

We have observed that the addition of T1w input does help in improving the quality of images in attempting to obtain an out of distribution target resolution. We observed a significantly better performance for our model that used a single upsampling rate of $1.6\times$. This indicates that the incorporation of a T1w image has been beneficial in the output quality at a previously unseen upsampling rate.

\begin{figure}[!t]
    \centering
    \includegraphics[width=0.7\textwidth]{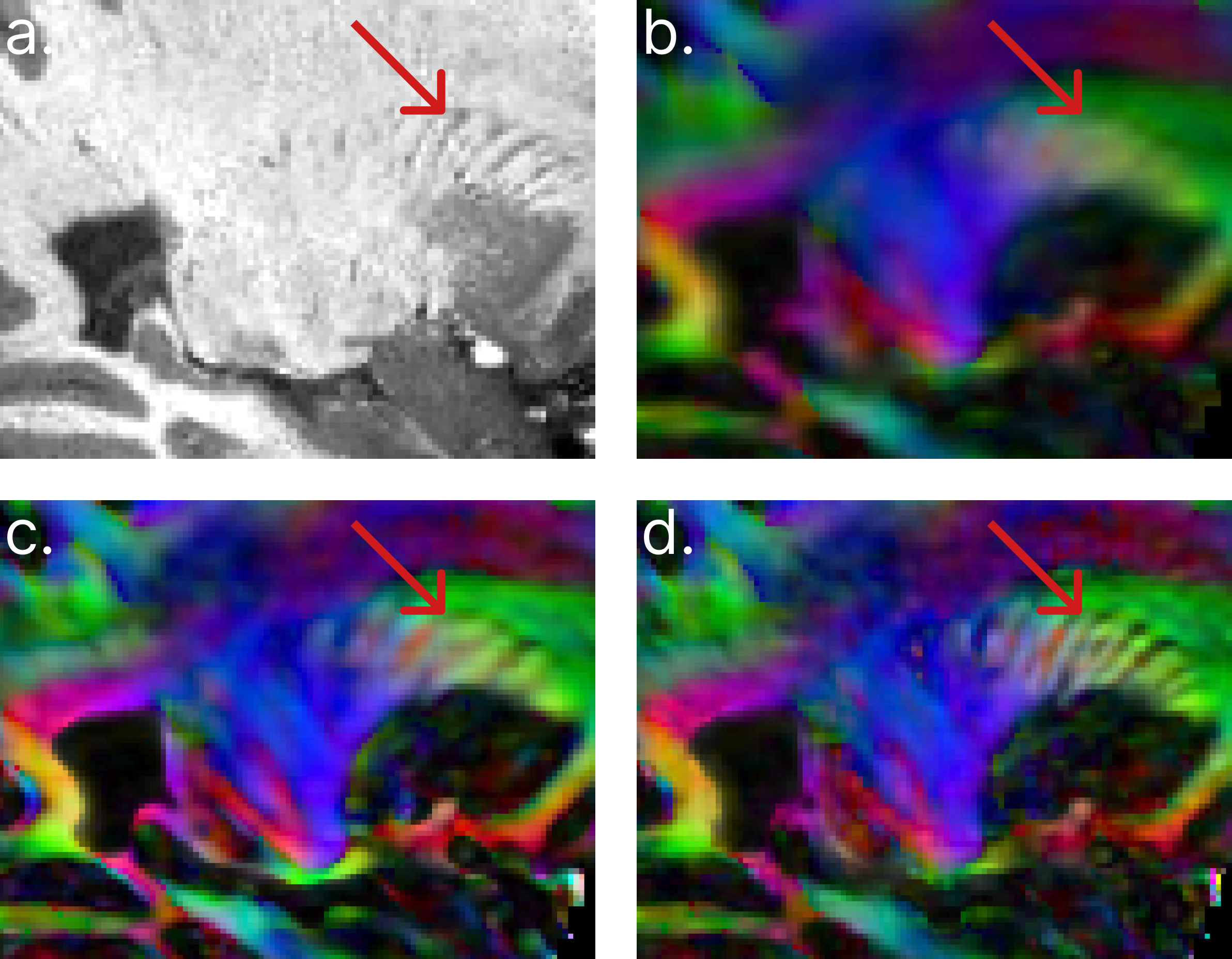}
    \caption{Qualitative comparison of T1w input with colour FA maps constructed from competing upsampling methods.
    We show corresponding sagittal slices from a 7T 0.7mm T1w companion volume (a), and Connectom 2~\si{mm} DTI upsampled using linear interpolation (b) a unimodal ablation model (c) and our multimodal network (d).
    We highlight striations in internal capsule which are visible in the T1w image and gain higher definition from our multimodal approach.}
    \label{fig:CFA-Qualitative}
\end{figure}

\autoref{fig:CFA-Qualitative} shows qualitative differences between the different model outputs on out-of-distribution images. We noticed certain features that resembled structures contained in the T1w image and not evident in the input dMRIs. In particular striations at fibre crossings between the Corpus Callosum and Corticospinal Tracts were sharper in the output of the multimodal model than in the ablation model.

\section{Discussion}

We have implemented an IQT model to upsample DTI data guided by high-resolution T1w images. Our tests showed that including the T1w input improved the output quality, especially in configurations where the DTI input had lower resolution and hence was less reliable. In addition, from our qualitative results we were able to view certain structures in our DTI output that were originally present in the T1w input and not evident in the DTI input.
However, we have also noticed that while the addition of the T1w image improved the upsampling quality at every test we conducted, we have observed diminishing returns at higher DTI input resolutions. This further indicates that our model essentially uses structural information from the T1w input to incorporate into the DTI data when there is a lack of said structural information in the DTI input.
Among the models including the T1w input, the ones trained on a single resolution closest to the test image performed the best in each experiment as it could be expected. The model trained with multiple resolutions, however, had a similar performance to the best one in each case and thus showed more robustness to different test cases.

To our knowledge, our approach is the first dMRI SR model that aims to combine multiple MRI modalities to reliably go beyond the resolution of its training set. Our approach, once qualitatively and quantitatively validated with the use of an ultra-high-resolution DTI scan, could be used in many applications to provide a reliable and quicker way to obtain high-resolution DTI scans. For example, transferring structural information of T1w scans from one scanner into diffusion scans acquired by a different scanner, as we demonstrated through our qualitative analysis. In addition to possibly improving the upsampling quality of dMRI scans, we speculate that the addition of a second MRI modality can be further used to reveal certain features that otherwise would not have been noticeable. For example, according to Wang et al.~\cite{10.3389/fneur.2021.591586}, certain brain lesions related to pharmacoresistant focal epilepsy such as Focal Cortical Dysplasia would be better detected in a 7T scanner because they are more clearly visible and have better-defined boundaries. By using high-resolution structural MRI images to upsample dMRI data, we our model could thus enhance the utility of dMRI for Focal Cortical Dysplasia. Similarly, the enhanced anatomical detail of the dMRI output of our model could find diagnostic or research applications in other medical conditions.

For future work, we are planning to develop our fusion model by modifying the network architecture and the training configuration to more effectively fuse features from the T1w input. Currently, our models are configured to upsample to one specific resolution that the T1w input is at. 
This configuration can therefore present a possibility where the model learns to upsample to the training target resolution rather than the T1w input resolution.
We are planning on mitigating this condition using multiple target resolutions, similar to the use of multiple resolutions on input. We are also planning on applying further data augmentations in the Diffusion Weighted Imaging (DWI)\index{Diffusion Weighted Imaging} sequence to further ensure the robustness of our model. Additionally, we are considering using different architectures that may fuse our data better such as by incorporating attention layers initially proposed by Vaswani et al.~\cite{vaswani2017attention}, or with the use of a diffusion model by Ho et al.~\cite{ho2020denoising}.
Finally, our model is currently implemented to work with T1w and DTI data. We will investigate the use of other structural modalities, such as T2-weighted or FLAIR images, that could provide different and potentially useful information. We are also planning to extend the method to multi-shell dMRI techniques, e.g. Diffusion Kurtosis Imaging~\cite{jensen2005diffusional} or Mean Apparent Propagator MRI~\cite{OZARSLAN201316}, that would allow a more advanced and specific charcterisation of brain microstructure.

To conclude, we have proposed an IQT model to combine information from a secondary MRI modality for upsampling dMRI models. Our proposed approach does demonstrate an increase in upsampling performance with the incorporation of a secondary MRI modality. Our plan is to develop this tool further by ensuring the dependence of our model on our T1w data with the use of different network architectures, or by eliminating possible ambiguity in our training configuration.

\begin{credits}
\subsubsection{\ackname} This study was funded by Wellcome Trust award 221915/Z/20/Z, MRC award MR/W031566/1 and was supported by the NIHR UCLH Biomedical Research Centre and the Medical Research Council (grant MR/W031566/1). Data were provided in part by the Human Connectome Project WU-Minn Consortium (Principal Investigators: David Van Essen and Kamil Ugurbil; 1U54MH091657), funded by the 16 NIH Institutes and Centers that support the NIH Blueprint for Neuroscience Research; and by the UK National Facility for In Vivo MR Imaging of Human Tissue Microstructure funded by the EPSRC (grant EP/M029778/1), and The Wolfson Foundation, and supported by a Wellcome Trust Strategic Award (104943/Z/14/Z).

\end{credits}
%
%
%
\bibliographystyle{splncs04}
\bibliography{references.bib}




\end{document}